\documentclass{sig-alternate-05-2015}

\usepackage{color}

\begin{document}
\toappear{}
\setcopyright{acmcopyright}

\hyphenation{dig-it-al meth-od-ol-o-gies}

\doi{10.475/123_4}

\isbn{123-4567-24-567/08/06}

\conferenceinfo{FSE '16}{November 13--18, 2016, Seattle, WA, USA}

\acmPrice{\$15.00}

\conferenceinfo{FSE}{'16, Seattle, WA, USA}

\title{Bounded Model Checking of State-Space Digital Systems}
\subtitle{The Impact of Finite Word-Length Effects on the Implementation of Fixed-Point Digital Controllers Based on State-Space Modeling}

\numberofauthors{1}

\author{
%
%
Felipe R. Monteiro\\
       \affaddr{Federal University of Amazonas}\\
       \affaddr{Manaus, Amazonas, Brazil}\\
       \email{felipemonteiro@ufam.edu.br}
}

\date{30 July 1999}

\maketitle
\begin{abstract}
The extensive use of digital controllers demands a growing effort to prevent design errors that appear due to finite-word length (FWL) effects. However, there is still a gap, regarding verification tools and methodologies to check implementation aspects of control systems. Thus, the present paper describes an approach, which employs bounded model checking (BMC) techniques, to verify fixed-point digital controllers represented by state-space equations. The experimental results demonstrate the sensitivity of such systems to FWL effects and the effectiveness of the proposed approach to detect them. To the best of my knowledge, this is the first contribution tackling formal verification through BMC of fixed-point state-space digital controllers.
\end{abstract}

%
%

\ccsdesc[500]{Computer systems organization~Real-time systems}
\ccsdesc[300]{Computer systems organization~Embedded systems}
\ccsdesc[500]{Software and its engineering~Model checking}
\ccsdesc[100]{Software and its engineering~Formal methods}
\ccsdesc[300]{Theory of computation~Verification by model checking}

\printccsdesc

\keywords{Real-time Systems; Model Checking; State-Space; Formal Verification; Digital Controllers.}

\section{Motivation}

In real-time systems, digital controllers are algorithms that manipulate digital signals, in order to influence the behavior of a system~\cite{Ogata2001}; it can be mathematically expressed as difference equations, transfer functions, or state-space equations. In this particular work, the focus is on state-space models, which represent the behavior of a system through a state evolution equation $\dot{x}(n+1)$ and an instantaneous output equation $y(n)$, as follows: 

\vspace{-5pt}
\begin{equation}
\begin{split}
\dot{x}(n+1) &= A x(n) + B u(n)
\\
y(n) &= C x(n) + D u(n), 
\end{split}\label{eq:ss-example}
\end{equation}

\noindent where $A$, $B$, $C$, and $D$ are matrices that fully specify a digital system. Such models can be translated into algorithms and implemented in several kinds of microprocessors ({\it e.g.}, field programmable gate arrays (FPGA) devices~\cite{4267891} and digital signal processors~\cite{MASTEN1997449}). Importantly, each one of these platforms can manipulate and represent numbers using different formats and arithmetics ({\it e.g.}, number of bits, fixed- or floating-point arithmetic), which can directly affect the performance and precision of the digital-control system~\cite{Bessa2016}. In fact, such systems are vulnerable to finite word-length (FWL) effects~\cite{Guang2013, Istepanian2001}, which can cause several quantization problems, such as truncation or round-off errors. Particularly, in such circumstances, the precision of each element from matrices $A$, $B$, $C$, and $D$ will be affected by FWL effects, which can compromise the system's properties ({\it e.g.}, stability). Additionally, fixed-point processors present high processing speed with reduced cost, which makes them a valuable choice for designing digital controllers; nonetheless, such an approach might lead to more nonlinearities, round-off errors, and overflows. 




In order to tackle such problem, this paper proposes a verification methodology based on bounded model checking (BMC) techniques~\cite{Clarke2009}, which verifies properties on state-space digital controllers, by means of a verification tool named as Digital-Systems Verifier (DSVerifier). It is worth noting that this paper extends a previous work~\cite{Bessa2016, Ismail2015, Abreu2016, 7091165, 7048514}. In particular, the major improvement of the DSVerifier version described here relies on the support for state-space models, which allows a better insight about the internal system behavior, enables the verification of new properties ({\it e.g.}, controllability and observability), and considers initial conditions for system analysis~\cite{fairman1998linear}. In addition, DSVerifier now supports two efficient model-checking tools as back-end: ESBMC~\cite{esbmc, Morse2014} (previously supported) and CBMC~\cite{CBMC, Clarke2004}.


\section{Background and Related Work}
\label{sec:background}

In order to deal with FWL effects on digital systems, some approaches suggest special metrics, search algorithms or methodologies to achieve an optimal word-length and avoid FWL effects~\cite{1104162, 4663846, 1219100, 781908, Mohta1998, 476465}. There are also simulation tools ({\it e.g.}, LabVIEW~\cite{Johnson1997} and MATLAB~\cite{Sigmon98a}), which are traditionally used by control engineers. However, such approaches depend on input stimulation to evaluate the state-space of a system, which might not exploit all possible conditions that a system can exhibit. In contrast, Alur {\it et al.}~\cite{113766, Alur1993} proposed the prior automated verification approaches, regarding model checking, which inspired the development of other verifiers for cyber-physical systems and hybrid automata ({\it e.g.}, Maellan~\cite{Synopsys2006}, Open-Kronos~\cite{Tripakis2005}, and UPPAAL~\cite{Behrmann2004}). Nonetheless, differently from the work presented here, such approaches do not tackle system robustness related to implementation aspects~\cite{Bessa2016, Ismail2015, Abreu2016}.


\section{Methodology}
\label{sec:approach}

\begin{figure}[h]
\centering
\includegraphics[width=.39\textwidth]{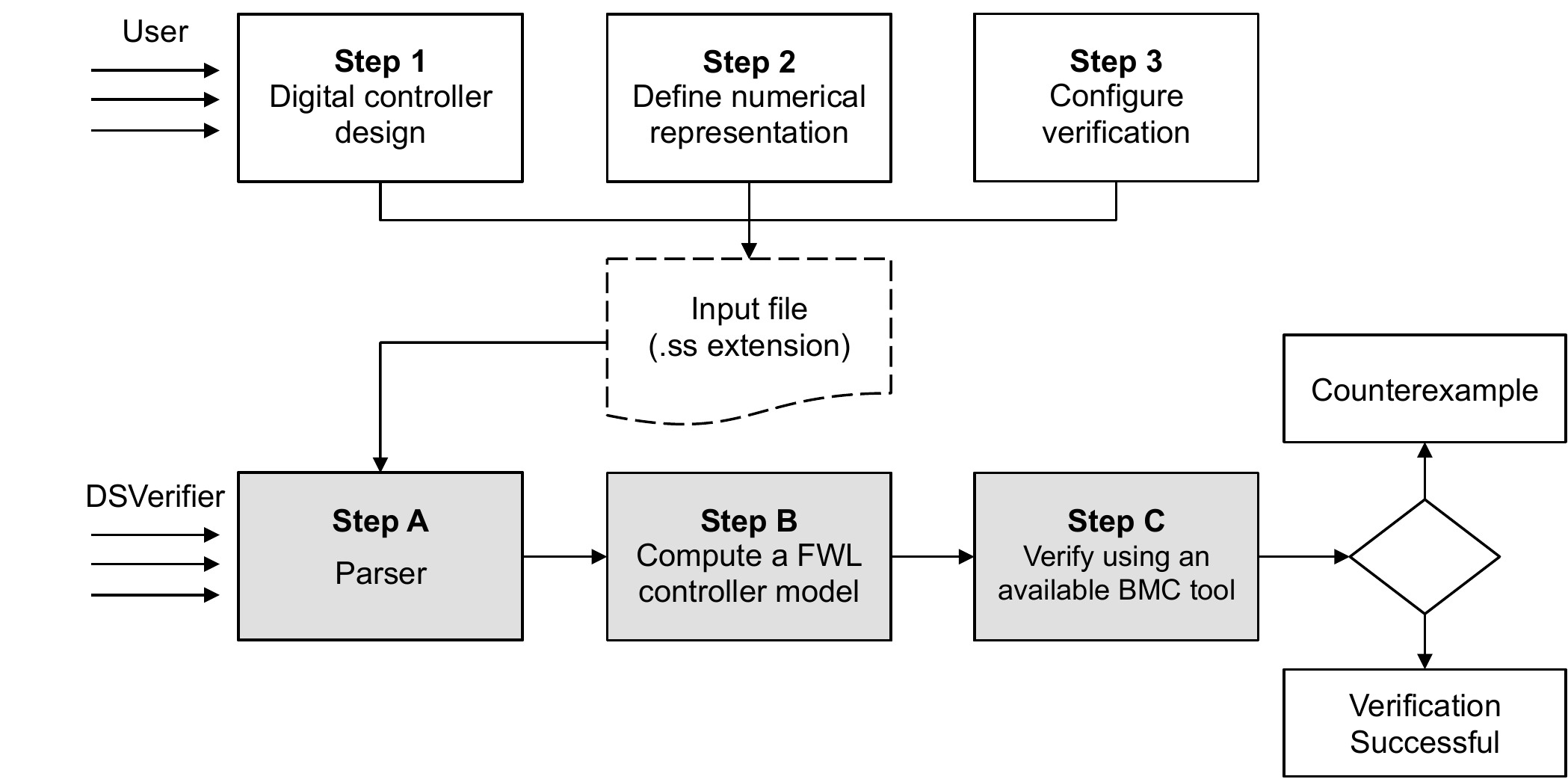}
\caption{Verification methodology.}
\label{dsverifier}
\end{figure}

DSVerifier works as front-end for BMC tools (with support to full ANSI-C verification), in order to verify state-space digital systems. As one can see in Figure~\ref{dsverifier}, the verification methodology proposed in this paper is split into two main stages as follows: manual (user) and automated (DSVerifier) procedures. In the former, the software engineer manually performs steps $1$ to $3$. Step $1$ is related to the design process of a digital system, while step $2$ to its implementation details, {\it i.e.}, numerical representation $<I,F>$, where $I$ is the number of bits for the integer part, and $F$ is the number of bits for the fractional part. Then, in step $3$ the user chooses a property $\phi$ to be verified ({\it e.g.}, {\it quantization\_error}), a maximum verification time, a bound $k$, and a BMC tool. Importantly, all specifications from the previous steps are detailed in an input file using the same syntax as MATLAB code standard.

After that, DSVerifier receives the respective input file and then performs the verification of the desired property $\phi$; it is worth noting that steps $A$ to $C$ are completely automatic. In step $A$, DSVerifier builds an intermediate ANSI-C code for the digital system implementation. Then, in Step $B$, it formulates a FWL model using a function $FWL[\cdot]:\mathbb{R}\rightarrow Q[\mathbb{R}]$, which applies the FWL effects to a state-space digital system, where $Q[\mathbb{R}]$ represents the quantized set of representable real numbers in the chosen implementation format. Finally in the step $C$, the translation of the resulted ANSI-C code ({\it i.e.}, the respective quantized state-space digital system) into SAT or SMT formulae is completed, by a highly efficient bounded model-checking tool ({\it e.g.}, ESBMC or CBMC)~\cite{esbmc, CBMC}. Here, DSVerifier symbolically checks a given property $\phi$ w.r.t. digital systems. If any violation is found, then DSVerifier reports a counterexample, which contains system inputs that lead to a failure. A successful verification result is reported if the system is safe w.r.t.  $\phi$ up to a bound~$k$.

As aforementioned, DSVerifier supports the verification of the following properties regarding quantized digital system: {\bf Quantization error -} it checks whether the output quantization is inside a tolerable bound; {\bf Stability -} it checks digital-system stability using the Eigen Library~\cite{eigenweb}; {\bf Controllability -} it checks whether a digital system $M$ is controllable, based on the rank of its controllability matrix; and {\bf Observability - } it checks whether a digital system $M$ is observable, based on the rank of its observability matrix.






It is worth noting that all numerical operations are performed through fixed-point arithmetic, according to a certain precision set by the user, and all properties are sound and complete. In addition, all aforementioned verifications can be performed in a closed-loop configuration.

\section{Preliminary Results}
\label{sec:results}



For the following evaluation, an automatic test-suite was developed, with $25$ digital systems\footnote{DSVerifier, all benchmarks, and a detailed test evaluation are available at www.dsverifier.org/} extracted from literature~\cite{Abdelzaher2008, Kuo1992}. In particular, this study employs CBMC~$v5.4$, with the SAT solver MiniSAT~$v2.2.0$~\cite{Cimatti2013}. All systems are checked against four properties, as described in Section~\ref{sec:approach}, using a 32-bits micro-controller hardware configuration with three precisions ($8$, $16$, and $32$-bits), which results in $300$ verifications. 

\begin{figure}[h]
\centering
\includegraphics[width=.45\textwidth]{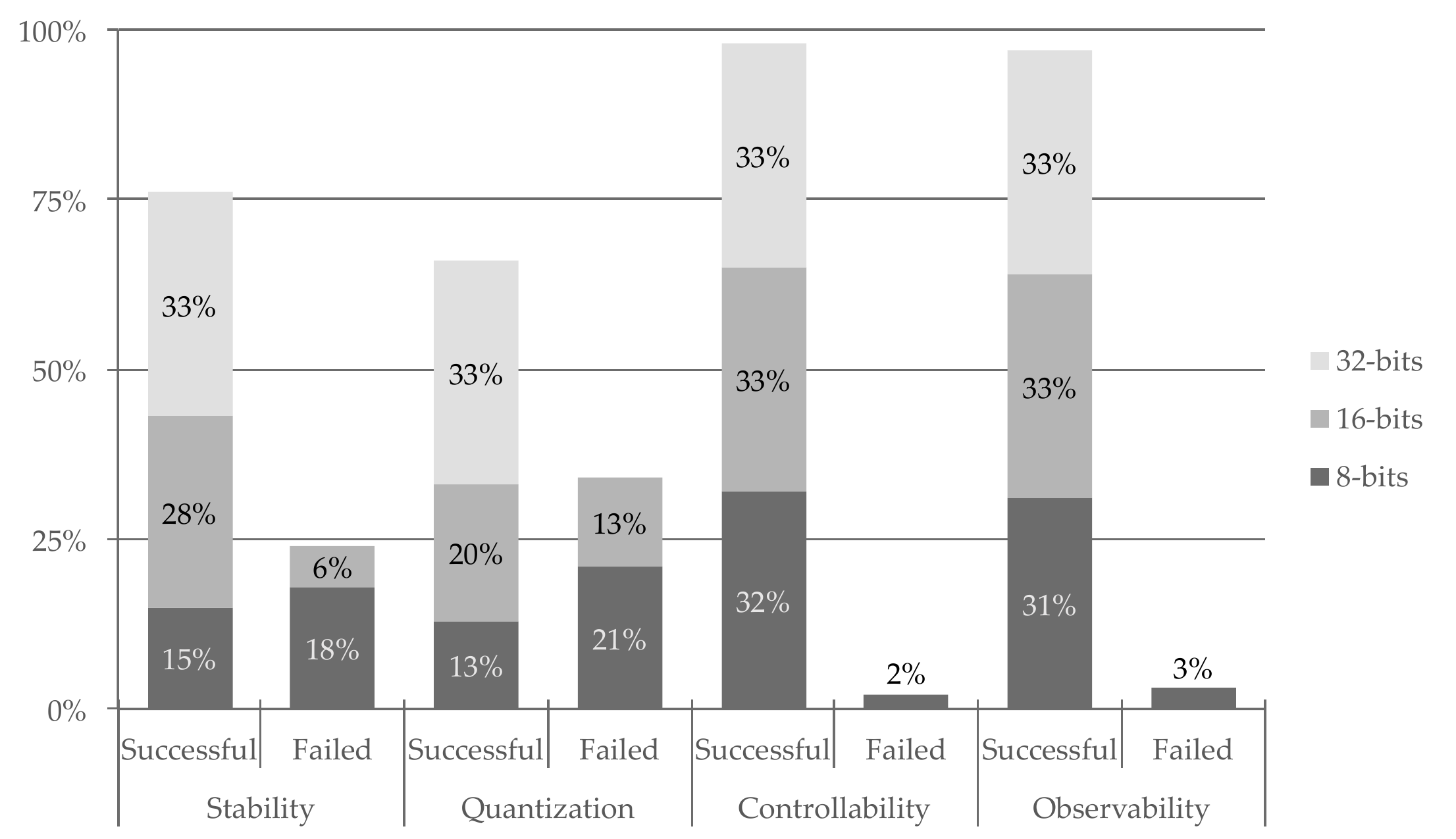}
\caption{Experimental results.}
\label{results}
\end{figure}

Indeed, all components of the test-suite are stable, controllable, and observable; however, based on the experimental results shown in Figure~\ref{results}, one may noticed that {\it (i)} the properties of a digital system might not be held, once quantization errors affect its representation, {\it (ii)} the lower the precision, the higher its sensibility to FWL effects, and {\it (iii)} controllability and observability are less sensitive to FWL effects, once they only rely on the system's coefficients. In addition, all $300$ verifications were performed in approximately $7$ hours. Finally, the failed cases were validated with Simulink~\cite{xue2013}, using the respective counterexample.


{\bf Contributions.} Particularly, this work makes four major contributions: {\bf (i)} support for state-space representations, {\bf (ii)} verification of quantization error for single-input and single-output (SISO) systems~\cite{Ogata2001}, {\bf (iii)} stability (for state-space systems), controllability and observability verifications for SISO and multi-input and multi-output (MIMO) systems~\cite{Ogata2001}, and {\bf (iv)} closed-loop verification for the aforementioned properties. To the best of my knowledge, this is the first report addressing formal verification through BMC of fixed-point digital controllers, based on the state-space representation. In future, other properties and BMC tools will be integrated into DSVerifier, in addition to support for systems with uncertainties.






\bibliographystyle{unsrt}
\bibliography{references}

\end{document}